\documentclass[prd,twocolumn]{revtex4}
\usepackage{dcolumn}
\usepackage{amsmath}
\usepackage{bm}

\begin{document}

%\preprint{Preprint Universit\'{e} de Mons-Hainaut}
%\draft
\title{About the Casimir scaling hypothesis}

\author{Claude \surname{Semay}}
\thanks{FNRS Research Associate}
\email[E-mail: ]{claude.semay@umh.ac.be}
\affiliation{Groupe de Physique Nucl\'{e}aire Th\'{e}orique,
Universit\'{e} de Mons-Hainaut,
Acad\'{e}mie universitaire Wallonie-Bruxelles,
Place du Parc 20, B-7000 Mons, Belgium}

\date{\today}

\begin{abstract}
A lattice calculation shows that the Casimir scaling hypothesis is well
verified in QCD, that is to say that the potential between two opposite
color charges in a color singlet is proportional to the value of the
quadratic Casimir operator. On the other hand, in a bag model
calculation for the same system, a scaling of the string tension with
the square root of the quadratic Casimir operator is obtained. It is
shown that, within the same formalism but with the assumption that the
width of the string is independent of the color charges, the string
tension is proportional to value of the quadratic Casimir operator. Some
considerations about the color behavior of the total interaction are
given.
\end{abstract}

% 12.39.Pn Potential models
% 12.39.Ba Bag model
% 12.38.Aw General properties of QCD (dynamics, confinement, etc.)

\maketitle

The Casimir scaling hypothesis means that the potential between two
opposite colour charges in a colour singlet is proportional to the value
of the quadratic Casimir operator. A lattice calculation \cite{bali00}
exclude any violations of this hypothesis that exceed 5\% for charge
separations of up to 1~fm. Nevertheless, other models do not predict
such a colour behaviour. For instance, a scaling of the string tension
with the square root of the quadratic Casimir operator is obtained in a
bag model calculation \cite{john76}. We show here that the Casimir
scaling can be obtained if the fundamental assumption in the bag model,
the existence of a confining pressure $B$, is replaced by the hypothesis
of the existence of a universal string section in the rest frame of the
charges.

We use the same formalism as in Ref.~\cite{john76}. Let us consider
two opposite colour charges with zero mass, moving attached by a string,
in a colour singlet. The colour electric flux $\vec E_a$ which leaves a
colour charge has the strength
\begin{equation}
\label{ea}
|\vec E_a| A = g \lambda_a,
\end{equation}
where $A$ is the cross section of the string, and $\lambda_a$ are the
colour matrices. If $x$ is the distance from the centre of mass (middle
of the string), a point of the string moves with the speed
\begin{equation}
\label{v}
v=\frac{2}{L} x,
\end{equation}
where $L$ is the length of the string. The colour charges at the
extremities move at the speed of light. The colour magnetic field, which
is produced by the rotation of the colour electric field, is given by
\begin{equation}
\label{ba}
\vec B_a = \vec v \times \vec E_a,
\end{equation}
at a point of the string which moves with velocity $\vec v$.
The quadratic Casimir operator $C$ is
\begin{equation}
\label{c}
C = \frac{1}{4}\sum_a \lambda_a^2.
\end{equation}

In Ref.~\cite{john76}, the section of the string is determined by the
surface equation of the bag containing the coloured particles. This
implies that its section $A$ is proportional to $\sqrt{C}$. In this
work, we assume that the section of the string is a constant $A_0$,
independent of $C$, in the rest frame of the string. Consequently, our
model is not a bag model, and no confining pressure $B$ is introduced.
When the string rotates, the section undergoes a Lorentz contraction
\begin{equation}
\label{av}
A = A_0 \sqrt{1-v^2}.
\end{equation}

To calculate the mass $M$ of the colour singlet system, let us first
compute the strength fields
\begin{equation}
\label{ea2}
\sum_a E_a^2 = \frac{4g^2C}{A^2} \quad \text{and} \quad
\sum_a B_a^2 = \frac{4g^2C}{A^2} v^2,
\end{equation}
the speed $\vec v$ of a point of the string being always perpendicular
to $\vec E_a$. All volume integrals are replaced by
\begin{equation}
\label{int}
\int d^3 x \rightarrow 2\int_{0}^{L/2} A dx = L \int_{0}^{1} A dv.
\end{equation}
The energy of the coloured flux lines is \cite{john76}
\begin{equation}
\label{ef1}
E_f = \frac{1}{2} \int d^3 x \sum_a \left( E_a^2 + B_a^2 \right).
\end{equation}
With the notations defined above, we obtain
\begin{equation}
\label{ef2}
E_f =2 g^2 C \frac{L}{A_0} \int_{0}^{1} \frac{1+v^2}{\sqrt{1-v^2}} dv.
\end{equation}
The angular momentum of the coloured flux lines is \cite{john76}
\begin{equation}
\label{jf1}
\vec J_f = \int d^3 x \sum_a \vec r \times \left( \vec E_a \times \vec
B_a \right).
\end{equation}
Thus, we obtain
\begin{equation}
\label{jf2}
J_f = 2 g^2 C \frac{L^2}{A_0} \int_{0}^{1} \frac{v^2}{\sqrt{1-v^2}} dv.
\end{equation}
Classically, a massless colour charge does not carry nor energy neither
momentum
\cite{laco89}. Consequently, the mass $M$ of the state is equal to $E_f$
and the total angular momentum $J$ is equal to $J_f$. We then obtain
\begin{equation}
\label{m2}
M^2 = \frac{9 \pi}{2} \frac{g^2}{A_0} C \, J = 18 \pi^2
\frac{\alpha_S}{A_0} C \, J,
\end{equation}
with $\alpha_S=g^2/4\pi$ the strong coupling constant. Let us note that,
in Ref.~\cite{john76}, the mass is determined from the condition
$\partial M/\partial L=0$. But this implies also that the contributions
of the massless colour charges to energy and momentum are vanishing.

We obtain the
linear Regge trajectories, but with a slope--that is to say a string
tension--proportional to $C$, and not to $\sqrt{C}$.
This result has already been obtained in Ref.~\cite{hans86}, but with a
different technique. With the more phenomenological approach used here,
we find that the energy density of the flux tube is given by
\begin{equation}
\label{mol}
\frac{M}{L} = 6 \pi^2 \frac{\alpha_S\, C}{A_0},
\end{equation}
which is quite different from the result of Ref.~\cite{hans86}.

In order to check the relevance of formula~(\ref{m2}), let us consider
the case of a meson, for which $C=4/3$. The relativistic flux tube model
\cite{laco89} predict that
\begin{equation}
\label{rft}
M^2 = 2\pi a \, J,
\end{equation}
where $a$ is the usual string tension. It is then possible to link the
section $A_0$ of the string to its tension $a$ and the strong coupling
constant $\alpha_S$
\begin{equation}
\label{a0}
A_0=12 \pi \frac{\alpha_s}{a}.
\end{equation}
The radius $R_0$ of the string is given by $\sqrt{A_0/\pi}$, assuming a
cylindrical form for the string.
For reasonable values of the QCD parameters, $\alpha_S \in [0.1-0.4]$
and $a \in [0.17-0.20]$~GeV$^2$, we find $R_0$ in the range 0.5-1.0~fm
\cite{hans86}.
A lattice calculation predicts a gaussian string width with a mean
radius around 0.35~fm \cite{bali95}. Given the simplicity of our model,
the agreement is quite reasonable.

We can expect that our model is relevant only if $L> 2R_0$. This
condition is satisfied if
\begin{equation}
\label{condj}
J > 8\pi C \alpha_s.
\end{equation}
Small values for $J$ are acceptable if the product $C \alpha_S$ is not
too large.

The key ingredient of this work is the assumption that the
width of the string is independent of the colour charges. Such a
possibility is also studied in recent works \cite{luci01,shos03}.
It could be interesting to test this hypothesis with lattice
calculations.

Besides the confinement, a one-gluon exchange process exists between the
two particles. The colour dependence of this interaction is given by
\begin{equation}
\label{oge}
\frac{1}{4}\sum_a \lambda_a(1) \lambda_a(2) =
\frac{1}{2} \left( 0-C-C \right) =-C.
\end{equation}
So we find again a colour scaling given by $C$. A constant potential
plays an important role in the hadron spectroscopy. In various
approaches \cite{grom81,simo01}, this constant is proportional to the
string tension. In this case the colour scaling is also given by $C$.
Finally, we can expect that the total potential between two opposite
colour charges in a colour singlet is proportional to the quadratic
Casimir operator, and not to its square root.

%The author thanks the FNRS (Fonds National de la Recherche
%Scientifique, Belgium) for its support.

%\appendix


\begin{thebibliography}{aa}
\bibitem{bali00} G.~S.~Bali, Phys. Rev. D {\bf 62}, 114503 (2000)
[hep-lat/0006022].
\bibitem{john76} K.~Johnson and C.~B.~Thorn, Phys. Rev. D {\bf 13},
1934 (1976).
\bibitem{laco89} D.~LaCourse and M.~G.~Olsson, Phys. Rev. D \textbf{39},
2751 (1989).
\bibitem {hans86} T.~H.~Hansson, Phys. Lett. B {\bf 166}, 343 (1986).
\bibitem{bali95} G.~S.~Bali, C.~Schlichter, and K.~Schilling, Phys. Rev.
D {\bf 51}, 5165 (1995) [hep-lat/9409005].
\bibitem {luci01} B. Lucini and M. Teper, Phys. Rev. D \textbf{64},
105019 (2001) [hep-lat/0107007].
\bibitem {shos03} A. I. Shoshi, F. D. Steffen, H. G. Dosch, and H. J.
Pirner, Phys. Rev. D \textbf{68}, 074004 (2003) [hep-ph/0211287].
\bibitem {grom81} D.~Gromes, Z. Phys. C {\bf 11}, 147 (1981); erratum
{\bf 14}, 94 (1982).
\bibitem{simo01} Yu.~A.~Simonov, Phys. Lett. B {\bf 515}, 137 (2001)
[hep-ph/0105141].
\end{thebibliography}
\end{document}